\documentclass[preprint,authoryear,12pt,natbib]{elsarticle}

\usepackage{epsfig}

\usepackage{amssymb}
\def\Mdot{\frac{dM}{dt}}
\def\Msol{\mbox{ }M_{\odot}}
\def\Lsol{\mbox{ }L_{\odot}}
\def\Rsol{\mbox{ }R_{\odot}}
\def\Tstar{\mbox{ }T_{\star}}
\def\Rstar{\mbox{ }R_{\star}}
\def\ergs{\mbox{ erg\,s}^{-1}}
\def\kms{\mbox{ km\,s}^{-1}}
\def\amin{^\prime}
\def\adeg{^{\circ}}
\def\nh{N_{\rm H}}
\def\cmmoinsdeux{\mbox{ cm}^{-2}}
\def\mags{\mbox{ magnitudes}}
\def\microns{\mbox{ } \mu \mbox{m}}
\def\igrjstu{\mbox{IGR~J16318$-$4848}}

\usepackage[ps2pdf,%
a4paper=true,%
breaklinks=true,%
colorlinks=true,%
pdfauthor={First Author et al.},%
pdftitle={Template for manuscripts in Advances in Space Research}%
]{hyperref}

\journal{Advances in Space Research}

\begin{document}

\begin{frontmatter}



\title{Optical/infrared observations unveiling the formation, nature and evolution of High-Mass X-ray Binaries\tnoteref{footnote1}}
\tnotetext[footnote1]{First, I would like to thank the organizers of the COSPAR event dedicated on SFXTs, Lara Sidoli and Vito Sguera. Then, I am grateful to the two anonymous referees who gave me useful comments and helped me to improve this review. Finally, I am lifelong indebted to my close collaborators on the study of {\it INTEGRAL} sources: A. Coleiro, P.A. Curran, Q.Z. Liu, I. Negueruela, L. Pellizza, F. Rahoui, J. Rodriguez, M. Servillat, J.A. Tomsick, J.Z. Yan, and particularly J.A. Zurita Heras for many fruitful discussions, and the realisation of Figures 1, 3 and 4. This work was supported by the Centre National d'Etudes Spatiales (CNES), based on observations obtained with MINE -- the Multi-wavelength {\it INTEGRAL} NEtwork --.}


\author{Sylvain Chaty\corref{cor}} 
\address{AIM (UMR-E 9005 CEA/DSM-CNRS-Universit\'e Paris Diderot) \\
Irfu/Service d'Astrophysique, Centre de Saclay,
FR-91191 Gif-sur-Yvette Cedex, France}
\address{Institut Universitaire de France,
103 boulevard Saint-Michel, 75005 Paris}
\cortext[cor]{Corresponding author}
\ead{chaty@cea.fr}


\begin{abstract}

In this review I first describe the nature of the three kinds of High-Mass X-ray Binaries (HMXBs), accreting through: {\it i.} Be circumstellar disc, {\it ii.} supergiant stellar wind, and {\it iii.} Roche lobe filling supergiants. 

I then report on the discovery of two new populations of HMXBs hosting supergiant stars, recently revealed by a wealth of new observations, coming from the high energy side ({\it INTEGRAL, Swift, XMM, Chandra} satellites), and complemented by multi-wavelength optical/infrared observations (mainly ESO facilities).
The first population is constituted of {\it obscured supergiant HMXBs}, the second one of {\it supergiant fast X-ray transients (SFXTs)}, exhibiting short and intense X-ray flares. 

I finally discuss the formation and evolution of HMXBs, constrain the accretion models (e.g. clumpy winds, transitory accretion disc, magneto-centrifugal barrier), show evidences suggesting the existence of an evolutionary link, include comparisons with population synthesis models, and finally build a consistent scenario explaining the various characteristics of these extreme celestial sources.

Because they are the likely progenitors of Luminous Blue Variables (LBVs), and also of neutron star/black hole binary mergers, related to short/hard gamma-ray bursts, the knowledge of the nature, formation and evolution of these HMXB populations is of prime importance.
\end{abstract}


\end{frontmatter}

\parindent=0.5 cm

\section{Introduction}

Nearly fifty years after the discovery of the first extra-solar X-ray source \citep[Sco\,X-1;][]{giacconi:1962}, X-ray astronomy has now reached its age of reason, with a plethora of telescopes and satellites covering the whole electromagnetic spectrum, obtaining precious observations of these powerful celestial objects populating our Universe.


High energy binary systems are composed of a compact object -- neutron star (NS) or black hole (BH) -- orbiting a companion star, and accreting matter from it \citep[cf e.g.][for a review]{chaty:2008a}. The companion star is either a low-mass star (typically $\sim 1 \Msol$ or less, with a spectral type later than B, called in the following LMXB for ``Low-Mass X-ray Binary''), or a luminous early spectral type OB high-mass companion star (typically $> 10 \Msol$, called HMXB for ``High-Mass X-ray Binary''). 300 high energy binary systems are known in our Galaxy: 187 LMXBs and 114 HMXBs \citep[respectively 62\% and 38\% of the total number;][]{liu:2006,liu:2007}.

Accretion of matter is different for both types of sources\footnote{We will not review here the so-called $\gamma$-ray binaries, where most of the energy is in the GeV-TeV range, coming from interaction between relativistic electrons of the pulsar and the wind of the massive star.}. In the case of LMXBs, the small and low-mass companion star fills and overflows its Roche lobe, therefore accretion of matter always occurs through the formation of an accretion disc. The compact object can be either a NS or a BH, Sco X-1 falling in the former category.
In the case of HMXBs, accretion can also occur through an accretion disc, for systems in which the companion star overflows its Roche lobe; however this is generally not the case, and there are two alternatives. The first one concerns stars with a circumstellar disc, and here it is when the compact object -- on a wide and eccentric orbit -- crosses this disc, that accretion periodically occurs (case of HMXBs containing a main sequence early spectral type Be III/IV/V star, rapidly rotating, called in the following BeHMXBs). The second case is when the massive star ejects a slow and dense wind radially outflowing from the equator, and the compact object directly accretes the stellar wind through e.g. Bondy-Hoyle-Littleton processes (case of HMXBs containing a supergiant I/II star, later called sgHMXBs).
We point out that there also exists binary systems for which the companion star possesses an intermediate mass (typically between 1 and $10 \Msol$, called IMXB for "Intermediate-Mass X-ray Binary").

We first describe in Section \ref{section:HMXBs} the various existing types of HMXBs, we then report in Section \ref{section:INTEGRAL} the new populations of HMXBs discovered by the {\it INTEGRAL} satellite, we give further details in Section \ref{section:formation-evolution} on the formation and evolution of HMXBs, and we finally conclude this review in Section \ref{section:conclusion}.

\section{Different types of High-Mass X-ray Binaries} \label{section:HMXBs}

\subsection{Be X-ray binaries}

BeHMXBs usually host a NS on a wide and eccentric orbit around an early spectral type B0-B2e (with emission lines) donor star, surrounded by a circumstellar ``decretion'' disc of gas created by a low velocity and high density stellar wind of $\sim 10^{-7} \Msol / yr$. This disc is characterized by an H$\alpha$ emission line (whose width is correlated with the disc size) and a continuum free-free or free-bound emission \citep[causing an infrared excess;][]{coe:2000, negueruela:2004}\footnote{For more details about BeHMXBs, we warmly recommend the excellent reference chapter by \cite{charles:2006} on optical/infrared emission of HMXBs.}.
They exhibit transient and bright X-ray outbursts: {\it i.} ``Type I'' are regular and periodic each time the NS crosses the decretion disc at periastron; {\it ii.} ``Type II'' are giant outbursts at any phase, with a dramatic expansion of the disc, enshrouding the NS; {\it iii.} ``Missed'' outbursts exhibit low H$\alpha$ emission (due to a small disc or a centrifugal inhibition of accretion), iv. ``Shifting outburst phases'' are likely due to the rotation of density structures in the circumstellar disc.

There are $\sim 50$ BeHMXBs in our Galaxy, and $>35$ in the Small Magellanic Cloud (SMC). This large number in the SMC, instead of $\sim 3$ predicted by the galactic mass ratio $\frac{Milky~Way}{SMC} \sim 50$, is likely due to the bridge of material between our Galaxy and the Magellanic Clouds \citep{mcbride:2008}. There is a large number of Supernova Remnants (SNRs) of similar age ($\sim 5$\,Myr) suggesting an increased stellar birth rate due to tidal interactions \citep{stanimirovic:1999}, since the previous closest SMC/Large Magellanic Cloud (LMC) approach was $\sim 100$\,Myr ago, and new massive stars probably formed the current HMXB population. 
A strong spatial correlation exists between emission line stars and stars aged $8-12$\,Myr, with BeHMXBs in the SMC \citep{meyssonnier:1993, maragoudaki:2001}, and the Magellanic Clouds conveniently provide an uniform sample of BeHMXBs in a compact region, all at the same distance. 
The number of HMXBs is therefore an indicator of star formation rate and starburst activity \citep{grimm:2003, bodaghee:2012b, coleiro:2013a}. 

\subsubsection{Formation of Be X-ray binaries}

BeHMXB populations are similar in the Milky Way and SMC/LMC, when examining the spectral type of the companion star \citep{negueruela:2002, mcbride:2008}. However, while comparing the spectral type of Be stars hosted in X-ray binaries versus isolated Be, all located in our Galaxy, it appears that the spectral type distribution of Be in X-ray binaries is much narrower, beginning at O9 V ($\sim 22 \Msol$), peaking at B0 V ($\sim 16 \Msol$) and stopping at B2 V ($\sim 8-10 \Msol$). The mass distribution of Be stars, ranging between 8 and $22 \Msol$, is consistent with the fact that wide orbits are vulnerable to disruption during SN event, especially for less massive B stars. The low-mass range therefore supports the existence of SN kick velocities and angular momentum loss, low-mass stars being ejected, while heavier systems become sgHMXBs \citep{portegies-zwart:1995, mcbride:2008}.

The formation of BeHMXBs has been explained by the so-called model of ``rejuvenation'' \citep{rappaport:1982}, by being a product of binary evolution. In this model, the mass transfer speeds up first the rotation of the outer layers of the B spectral type secondary star, and then the star itself, which starts to turn so quickly that a circumstellar disc of gas is created, giving birth to the Be phenomenon \citep[i.e. these stars are neither born as fast rotators, nor spun-up in the final main-sequence stages;][]{coe:2000, charles:2006}.
These systems are thus the result of moderately massive binaries undergoing a semi-conservative mass transfer evolution, with wide orbits (200 to 600\,days) produced before the first SN event of the system, and eccentric orbits due to small asymmetries during the SN event \citep{vandenheuvel:1983, verbunt:1995}. This model is consistent with the spectral types of the companion stars between O9 and B2V: stars less massive than $\sim 8 \Msol$ are ejected during the SN event, and stars more massive than $\sim 22 \Msol$ become supergiant stars.

\subsubsection{Structure of Be circumstellar disc}

In BeHMXBs, Be circumstellar discs are naturally truncated due to the presence of the NS, as a result of tidal torques at certain resonance points (where the Keplerian period $P_K$ is an integer fraction of the orbital period $P_\mathrm{orb}$), thus preventing any accretion of matter beyond these points. 
The accretion on the NS is unlikely for BeHMXBs where the NS is on a circular orbit, with a disc always truncated at a fixed size, smaller than the Roche lobe. This creates a persistent low-level X-ray emission from the stellar wind, and occasional Type II outbursts.
Instead, BeHMXBs with a NS on a highly eccentric orbit will allow for periodic accretion, with the size of the rotating truncated disc depending on the orbital phase, and at periastron the disc can include the NS orbit, thus creating Type I X-ray outbursts \citep{okazaki:2001, reig:2007}.

In addition, recent observations show that circumstellar discs in BeHMXBs also exhibit cycles of activity, likely due to their formation and dispersion. BeHMXBs therefore exhibit 3 periods: a spin period $P_\mathrm{spin}$, an orbital period $P_\mathrm{orb}$ and a super-orbital period $P_\mathrm{sup}$.
The first BeHMXB to have shown this cycle of activity is A\,0538-66, with $P_\mathrm{orb} = 16.65$\,days and $P_\mathrm{sup} = 421$\,days \citep{mcgowan:2003}.
Then, a comprehensive study of the MACHO and OGLE light curves over 18\,years of $\sim 20$ BeHMXBs show that $P_\mathrm{sup}$, extending from 500 to 4000\,days, seems to be correlated to $P_\mathrm{orb}$, from 10 to 300\,days \citep{rajoelimanana:2011}. While the origin of this variation is still unknown, this correlation shows that it is certainly due to binarity, and not intrinsic to the Be star.

\subsection{Supergiant X-ray binaries}

These systems usually host a NS on a circular orbit around an early spectral type supergiant OB donor star, with a steady wind outflow. They are separated in two distinct groups: Roche lobe overflow and wind-fed systems\footnote{It is interesting to note that Cyg\,X-1 is the only sgHMXB with Roche lobe overflow and stellar wind accretion hosting a confirmed BH, and therefore probably a rare product of stellar evolution in X-ray binary systems \citep[see e.g.][]{podsiadlowski:2003}.}. 
The former group constitutes the classical «bright» sgHMXBs with accreted matter flowing via inner Lagrangian point to the accretion disc, causing a high X-ray luminosity ($L_X \sim 10^{38} \ergs$) during outbursts.
The later group concerns close systems ($P_\mathrm{orb} < 15$\,days) with a low eccentricity, the NS accreting from deep inside the strong steady radiative and highly supersonic stellar wind. These systems exhibit a persistent X-ray emission at regular low-level effect ($L_X \sim 10^{35-36} \ergs$), on which are superimposed large variations on short timescales, due to wind inhomogeneities.
During their long term evolution, the orbits of sgHMXBs will tend to circularize more rapidly with time, while the rate of mass transfer steadily increases \citep{kaper:2004}.

A milestone in the evolution of these close binary systems takes place during the so-called ``common envelope phase''. This phase is initiated when the compact object penetrates inside the envelope of the companion star, in a rapidly decreasing orbit due to a large loss of orbital angular momentum. This phase has been invoked by \cite{paczynski:1976} to explain how high energy binary systems with very short $P_\mathrm{orb}$ can be formed, while both components of these systems -- large stars at their formation -- would not have been able to fit inside a binary system with such a small orbital separation. This phase of inward spiralling, currently taken into account in population synthesis models, but never observed yet, probably because it is short \cite[models predict a maximum duration of common envelope phase of only $\sim 1000$\,years;][]{meurs:1989} compared to the lifetime of a massive star ($\sim 10^{6-7}$\,years), is an ingredient of prime importance to understand the evolution of high energy binary systems \citep{tauris:2006}.

\subsection{The Corbet diagram: a tool to distinguish Be and sgHMXBs} \label{section:corbet}

\begin{figure}
\centerline{\includegraphics[width=6.6cm]{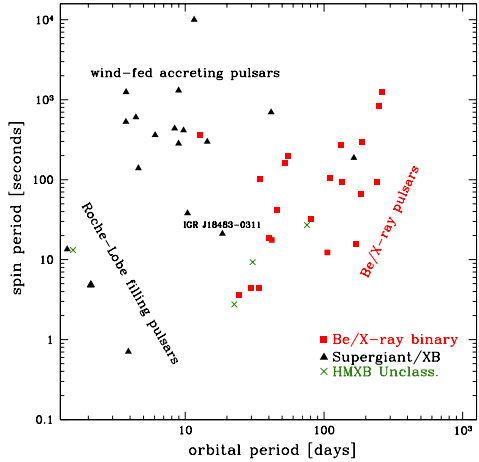}}
\caption{\label{corbet} Corbet diagram (NS $P_\mathrm{spin}$ versus system $P_\mathrm{orb}$), showing the three HMXB populations: {\it i.} systems hosting Be stars (BeHMXBs), {\it ii.} supergiant stars overflowing their Roche lobe, and {\it iii.} systems accreting the stellar wind of supergiant stars (sgHMXBs). The distinct location of these three types of HMXBs results from the interaction and equilibrium reached between accretion of matter and NS spin. 
The ``misplaced'' SFXT IGR\,J18483-0311 is indicated (see Section \ref{section:IGRJ18483}; credit J.A. Zurita Heras).
}
\end{figure}

In case of an accreting NS in an HMXB, it is possible to detect a pulsation, corresponding to the NS spin period $P_\mathrm{spin}$. These X-ray accretion-powered pulsars are divided according to their nature and the main accretion process -- BeHMXBs (decretion disc-accreting systems) or sgHMXBs (wind-accreting and Roche lobe overflow systems) -- in distinct locations in the Corbet diagram, representing the NS $P_\mathrm{spin}$ versus the system $P_\mathrm{orb}$ \cite[][see Figure \ref{corbet}]{corbet:1986}. This diagram is a valuable tool to study the interaction and feedback between the NS and accreted matter, the location of the different systems being determined by the equilibrium period reached by the rotation of the NS accreting matter on its surface.

It is clear from this diagram that a strong correlation exists for BeHMXBs -- $P_\mathrm{spin}\propto (P_\mathrm{orb})^2$ \citep{corbet:1984} --, due to the efficient transfer of angular momentum when the NS accretes material from the Be star decretion disc. Intuitively, a small (wide) orbit presents on average a high (low) stellar wind density, increasing (decreasing) the accretion pressure, and thus accelerating (slowing) the NS rotation, which in turn increases (decreases) the centrifugal inhibition, preventing (allowing) more accretion of matter. 
In contrast, the lack of correlation for sgHMXBs suggests that in this case, wind accretion is very inefficient to transfer angular momentum.

High energy properties, and in particular the accretion efficiency, of a compact object accreting from a circumstellar disc, a dust cocoon or the stellar wind of the companion star, will strongly differ, depending on the geometry of the system, altogether with the size and shape of the optically thin and fully ionized region of gas, similar to the Str\"omgren sphere \cite[see e.g.][]{pringle:1973, hatchett:1977}.
In particular, $P_\mathrm{spin}$ in HMXBs is determined by the stellar wind characteristics. sgHMXBs exhibit a spherically-symmetric radiation-driven wind, with a density $\rho \propto r^{-2}$ and a velocity $\sim 600-900\kms$, while BeHMXBs present an equatorial stellar wind with a density $\rho \propto r^{-3 \mbox{ to } -3.5}$ dropping faster, and a lower velocity $\sim 200-300 \kms$ \citep{waters:1988}. Therefore, as stated in \citet{waters:1989}, larger density and velocity gradients --at the distance of the NS-- allow a wind-fed accretion transfer of angular momentum which is by nature more efficient in BeHMXBs than sgHMXBs.

Accretion can occur on a magnetized NS only if the pressure of the infalling material is greater than the centrifugal inhibition (corresponding to the Alfven radius located inside the magnetospheric boundary). The NS should then reach an equilibrium rotation period $P_\mathrm{eq}$ corresponding to the corotation velocity $V_C$ (at the magnetospheric radius) equal to the Keplerian speed $V_K$. If $V_C > V_K$ (corresponding to $P_\mathrm{spin} < P_\mathrm{eq}$) the Propeller mechanism increases $P_\mathrm{spin}$ by ejecting material, taking away angular momentum \citep{illarionov:1975}. If $V_C < V_K$ (corresponding to $P_\mathrm{spin} > P_\mathrm{eq}$) the accretion either reduces or increases $P_\mathrm{spin}$ (spinning up or down), depending on the direction of angular momentum with respect to the NS spin. 

Taking into account the stellar wind density and assuming a steady accretion rate with the angular momentum of same direction than the rotating NS, $P_\mathrm{spin}$ should reach $P_\mathrm{eq} \propto \rho^{-3/7}$ \citep{waters:1988}. However, as pointed out by \citet{king:1991, waters:1989}, this is not the case, neither for BeHMXBs nor for sgHMXBs: $P_\mathrm{spin}$ of BeHMXBs randomly changes with the wind, and $P_\mathrm{spin}$ of sgHMXBs, higher than $P_\mathrm{eq}$, is typical of a NS embedded in the stellar wind of an O-type main sequence star.

\section{The $\gamma$-ray sky seen by the {\it INTEGRAL} observatory} \label{section:INTEGRAL} 
The {\it INTEGRAL} observatory is an ESA satellite launched on 17 October
2002 by a PROTON rocket on an eccentric orbit. It hosts 4
instruments: 2 $\gamma$-ray coded-mask telescopes -- the imager IBIS
and the spectro-imager SPI, observing in the range 10 keV-10 MeV, with
a resolution of $12\amin$ and a field-of-view of $19\adeg$ --, a
coded-mask telescope JEM-X (3-100 keV), and an optical telescope
(OMC).
%
%
The $\gamma$-ray sky seen by {\it INTEGRAL} is very rich, since 723 sources
have been detected by {\it INTEGRAL}, reported in the $4^{th}$ IBIS/ISGRI soft
$\gamma$-ray catalogue, spanning nearly seven years of observations in the 17-100
keV domain \citep{bird:2010}\footnote{See an up-to-date list at {\em http://irfu.cea.fr/Sap/IGR-Sources}, maintained by J. Rodriguez and A. Bodaghee}.  
%
Among these sources, there are 185 X-ray binaries (representing 26\% of the whole sample of sources detected by {\it INTEGRAL}, called ``IGRs'' in the following), 255 Active Galactic Nuclei (35\%), 35 Cataclysmic Variables (5\%), and $\sim 30$ sources of other type (4\%): 15 SNRs, 4 Globular Clusters, 3 Soft $\gamma$-ray Repeaters, 2 $\gamma$-ray bursts, etc. 
215 objects still remain unidentified (30\%).
X-ray binaries are separated into 95 LMXBs and 90 HMXBs, each category representing $\sim 13$\% of IGRs. Among identified HMXBs, there are 24 BeHMXBs and 19 sgHMXBs, representing respectively 31\% and 24\% of HMXBs.

It is interesting to follow the evolution of the ratio between BeHMXBs and sgHMXBs \cite[see e.g.][]{chaty:2011b}. During the pre-{\it INTEGRAL} era, HMXBs were mostly BeHMXBs. For instance, in the catalogue of 130 HMXBs by \citet{liu:2000}, there were 54 BeHMXBs and 5 sgHMXBs (respectively 42\% and 4\% of the total number of HMXBs). Then, the situation changed drastically with the first HMXBs identified by {\it INTEGRAL}: in the catalogue of 114 HMXBs (+128 in the Magellanic Clouds) of \citet{liu:2006}, 60\% of the total number of HMXBs were firmly identified as BeHMXBs and 32\% as sgHMXBs. Therefore, while the BeHMXB/HMXB ratio increased by a factor of 1.5, the sgHMXB/HMXB ratio increased by a larger factor of 8.
The ISGRI energy range ($> 20$\,keV), immune to the absorption that prevented the discovery of intrinsically absorbed sources by earlier soft X-ray telescopes, allowed us to go from a study of individual sgHMXBs (such as GX\,301-2, 4U\,1700-377, Vela\,X-1, etc.) to a comprehensive study of the characteristics of a whole population of HMXBs \cite[see for instance][and section \ref{section:formation} on the Galactic distribution of HMXBs]{coleiro:2013a}.




The most important result of {\it INTEGRAL} to date is the discovery of many new high energy sources -- concentrated in the Galactic plane, mainly towards tangential directions of Galactic arms, rich in star forming regions --, exhibiting common characteristics which previously had rarely been seen \cite[see e.g.][]{chaty:2008, coleiro:2013b}. Many of them are HMXBs hosting a NS orbiting an OB companion, in most cases a supergiant star. Nearly all the {\it INTEGRAL} HMXBs for which both $P_\mathrm{spin}$ and $P_\mathrm{orb}$ have been measured are located in the upper part of the Corbet diagram (see Figure \ref{corbet}). They are X-ray wind-accreting pulsars typical of sgHMXBs, with longer pulsation periods and higher absorption (by a factor $\sim4$) compared to previously known sgHMXBs \cite[see e.g.][]{bodaghee:2007}.
They divide into two classes: some are very obscured, exhibiting a huge intrinsic and local extinction, -- the most extreme example being the highly absorbed source IGR~J16318-4848 \citep{filliatre:2004} --, and the others are sgHMXBs exhibiting fast and transient outbursts -- an unusual characteristic among HMXBs --.  These are therefore called Supergiant Fast X-ray Transients \cite[SFXTs, ][]{negueruela:2006a, sguera:2005}, with IGR~J17544-2619 being their archetype \citep{pellizza:2006}.

\subsection{Obscured HMXBs: the example of IGR~J16318-4848}

  IGR~J16318-4848 was the first source discovered by IBIS/ISGRI on
  {\it INTEGRAL} on 29 January 2003 \citep{courvoisier:2003}, with a
  $2 \amin$ uncertainty.  {\it XMM-Newton} observations revealed a
  comptonised spectrum exhibiting an unusually high level of
  absorption: $\nh \sim 1.84 \times 10^{24} \cmmoinsdeux$
  \citep{matt:2003}.  {\it XMM-Newton} accurate localisation 
  allowed \citet{filliatre:2004} to rapidly trigger ToO photometric and
  spectroscopic observations in optical/infrared, leading to the
  confirmation of the optical counterpart \citep{walter:2003} and to
  the discovery of a bright infrared source \citep[B\,$>25.4\pm1$; I\,$=16.05\pm0.54$; J\,$= 10.33\pm 0.14$; H\,$=8.33\pm 0.10$ and K$_{\mathrm s} =7.20 \pm 0.05 \mags$;][]{filliatre:2004}. This source exhibits an
  unusually strong intrinsic absorption in the optical ($A_{\mathrm V} = 17.4
  \mags$), 100 times stronger than the interstellar absorption along
  the line of sight ($A_{\mathrm V} = 11.4 \mags$), but still 100 times lower
  than the absorption in X-rays.  
This led \citet{filliatre:2004} to
  suggest that the material absorbing in X-rays was concentrated
  around the compact object, while the one absorbing in
  optical/infrared was enshrouding the whole system.  

Near-infrared spectroscopy in the $0.95-2.5 \microns$ domain allowed \cite{filliatre:2004} to identify the nature of the companion star, by revealing an unusual spectrum, with many strong emission lines:
%
%
H and He~I (P-Cyg) lines characteristic of dense/ionised wind at $v = 400$\,km/s,
He~II lines signatures of a highly excited region,
$[$Fe~II$]$ lines reminiscent of shock heated matter,
Fe~II lines emanating from media of densities $>10^5-10^6$\,cm$^{-3}$, and
Na~I lines coming from cold/dense regions.
%
%
All these intense emission lines originate from a highly complex, stratified
circumstellar environment of various densities and temperatures,
suggesting the presence of an envelope and strong stellar outflow/wind
responsible for the absorption. Only luminous early spectral type stars show such extreme environments, and \citet{filliatre:2004} concluded that IGR~J16318-4848 was an unusual HMXB hosting a sgB[e] with characteristic luminosity $10^6 \Lsol$, mass $30 \Msol$, radius $20 \Rsol$ and temperature $T=20250$\,K, located at a distance between 1 and 6 kpc (see also \citeauthor{chaty:2005a} \citeyear{chaty:2005a}).
This source is therefore the second HMXB hosting a sgB[e] star,
after CI Cam \citep{clark:1999}.
Subsequently, \citet{rahoui:2008a} showed, by combining and fitting optical to mid-infrared (MIR) observations with a simple spherical blackbody, that IGR~J16318-4848 exhibited a MIR excess, 
interpreted as being due to the strong stellar outflow emanating from the sgB[e] companion star.  They found that this star had a temperature of $\Tstar=22000$\,K and radius $\Rstar = 20.4 \Rsol = 0.1$\,a.u., consistent with a supergiant star, with an extra component of temperature T $=1100$\,K and radius R\,$= 11.9\Rstar = 1$\,a.u., with A$_{\mathrm V} = 17.6 \mags$. 

\begin{figure}
  \centerline{\includegraphics[width=11.3cm]{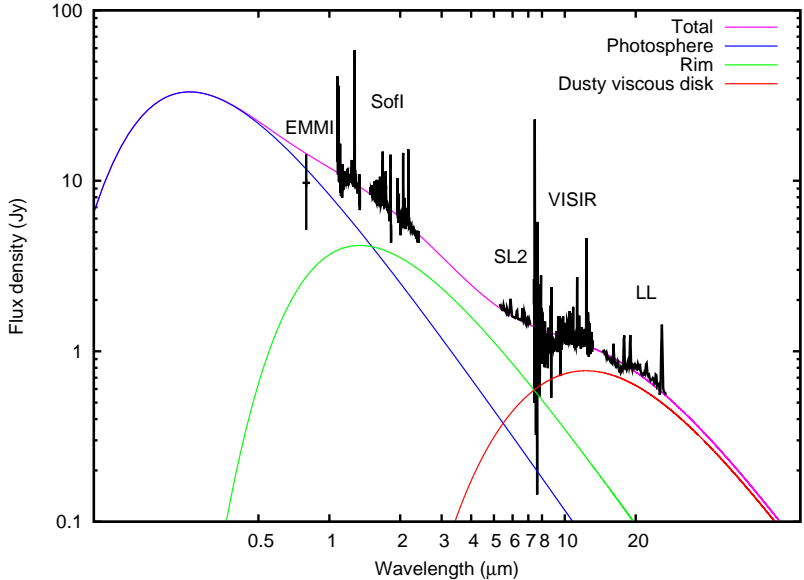}}
\caption{\label{figure:16318}
  Broadband near- to mid-infrared ESO/NTT+VISIR and {\it Spitzer} reddened spectrum of $\igrjstu$, from 0.9 to 35 $\mu$m, with fitted contribution of photosphere, rim and  dusty viscous disc \citep[from][]{chaty:2012}.
}
\end{figure}

\begin{figure}
  \centerline{\includegraphics[width=8.4cm]{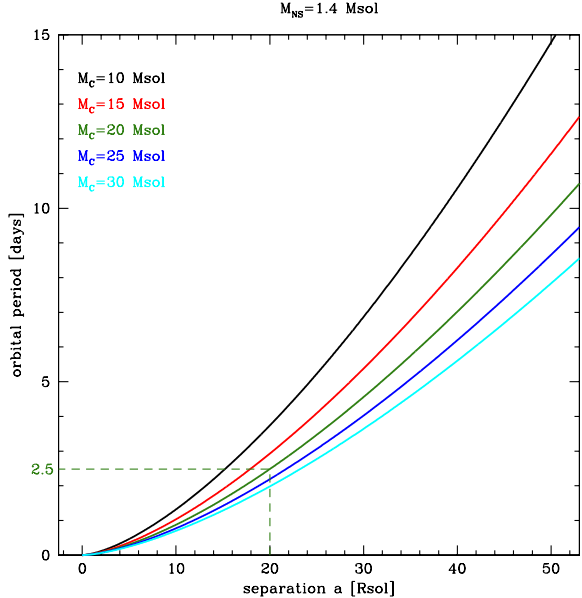}}
\caption{\label{figure:Porb-sep}
  Orbital period versus separation for HMXBs constituted of a NS orbiting a companion star of various masses (credit J.A. Zurita Heras).
}
\end{figure}

Recent MIR spectroscopic observations, with VISIR at the VLT and {\it Spitzer}, showed that the source was exhibiting strong emission lines of H, He, Ne, PAH, Si, proving that the extra absorbing component was made of dust and cold gas \citep{chaty:2012}. By fitting the optical to MIR spectra with a more sophisticated aspheric disc model developped for HAeBe objets, and adapted to sgB[e] stars, they showed that the supergiant star was surrounded by a hot rim of dust at $5500$\,K, with a warm dust component at $900$\,K around it \citep[see Figure \ref{figure:16318};][]{chaty:2012}.
By assuming a typical $P_\mathrm{orb}$ of 10\,days, we obtain an orbital separation of $\sim 50 \Rsol$ (see Figure \ref{figure:Porb-sep}), smaller than the extension of the extra component of dust/gas ($= 240 \Rsol$), suggesting that this dense and absorbing circumstellar material envelope enshrouds the whole binary system, like a cocoon of dust (see Figure \ref{figure:obscured-sfxt}, left panel).
We point out that this source exhibits such extreme characteristics that it might not be fully representative of all obscured sgHMXBs.

\subsection{Supergiant Fast X-ray Transients}


SFXTs constitute a new class of $\sim 15$ sources identified among the recently discovered IGRs. They are HMXBs hosting NS orbiting OB-spectral type supergiant stars, exhibiting peculiar characteristics compared to ``classical'' sgHMXBs: rapid outbursts --constituted of many fast individual flares-- rising in tens of minutes, typically lasting for $\sim 3$\,ks, and hard X-ray spectra requiring a BH or NS accretor. The frequency of these outbursts is $\sim7$\,days, their luminosity $L_X$ reaching $\sim 10^{36} \ergs$ at the peak. Inbetween these outbursts, there is a faint quiescent emission \citep[see e.g.][]{walter:2007}.

SFXTs can be divided in two groups, according to the duration and frequency of their outbursts, and their variability factor, defined as $\frac{L_\mathrm{max}}{L_\mathrm{min}}$ ratio \cite[typical hard X-ray flare-to-deep quiescence count rate, as in][]{walter:2007}:

\begin{enumerate}

\item ``Classical'' SFXTs exhibit a very low quiescence $L_X$ and a variability factor higher than 100, reaching $10^4$ for some sources, such as XTE\,J1739-302 and IGR\,J17544-2619, which exhibit flares in a few minutes \cite[see e.g. ][]{pellizza:2006}.

\item ``Intermediate'' SFXTs exhibit a higher $<L_X>$ and a lower variability factor, with longer flares. SFXTs might appear like persistent sgHMXBs with $<L_X>$ below the canonical value of $\sim 10^{36} \ergs$, and superimposed flares \citep{walter:2007}, but there might be some observational bias in these general characteristics, therefore the distinction between SFXTs and sgHMXBs is not yet well defined \cite[see the discussion on "intermediateness" in][]{smith:2012}.

\end{enumerate}


Such sharp rises exhibited by SFXTs are incompatible with the orbital motion of a compact object through a smooth medium \citep{negueruela:2006a, sguera:2005, smith:2006}. Instead, flares must be created by the interaction of the accreting compact object with the dense clumpy stellar wind (representing a large fraction of stellar $\Mdot$).  In this case, the flare frequency depends on the system geometry, and the quiescent emission is due to accretion onto the compact object of diluted inter-clump medium, explaining the very low quiescence level in classical SFXTs. We will now describe in more detail this mechanism, called the macro-clumping scenario.

\subsubsection{Macro-clumping scenario}

In this scenario, it is assumed that each SFXT flare is due to the accretion of a single clump, in the Bondi-Hoyle-Lyttleton approximation. Thus, the strong variability in the X-ray light curve directly results from the strong density and velocity fluctuations in the wind \cite[][]{oskinova:2012}.

The typical parameters in this scenario are: a compact object with large orbital radius: $10 \Rstar$, a clump size of a few tenths of $\Rstar$, a clump mass of $10^{19-22}g$ (for $\nh=10^{22-23}\cmmoinsdeux$), a mass loss rate of $10^{-(5-6)} \Msol/yr$, a clump separation of order $R_{\star}$ at the orbital radius, and a volume filling factor: 0.02$->$0.1 \cite[see e.g.][]{oskinova:2012}.
The flare to quiescent count rate ratio is directly related to the $\frac{clump}{inter-clump}$ density ratio, which ranges between 15-50 for intermediate SFXTs, and $10^{2-4}$ for classical SFXTs. 
A very high degree of porosity (macroclumping) is required to reproduce the observed outburst frequency in SFXTs, in good agreement with UV line profiles and line-driven instabilities at large radii \citep{runacres:2005, owocki:2007, walter:2007}. 
The number of clumps vs radius in a ring of given width and height, for a velocity law with beta=0.8 and porosity parameter $L_0=0.35$, is given in \cite{oskinova:2007}. 


To explain the emission of SFXTs in the context of sgHMXBs, \citet{negueruela:2008} invoked the existence of two zones in the stellar wind created by the supergiant star, of high and low clump density. This would naturally explain {\it i.} the X-ray light curves, each flare being due to the accretion of single clumps, several different flares adding up together to make an SFXT outburst, {\it ii.} the smooth transition between sgHMXBs and SFXTs, and {\it iii.} the existence of intermediate systems; the main difference between classical sgHMXBs and SFXTs being in this scenario the NS orbital radius.
Indeed, a basic model of porous wind with macro-clumping \citep{negueruela:2008} predicts a substantial change in the properties of the wind ``seen by the NS'' at a distance $r \sim 2 \Rstar$, where we stop seeing persistent X-ray sources. There are 2-regimes:
at $r < 2 \Rstar$ the NS sees a large number of clumps, embedded in a quasi-continuous wind; and at $r > 2 \Rstar$ the density of clumps is so low that the NS is effectively orbiting in an empty space.
NS in classical sgHMXBs can therefore only lie within $2 \Rstar$ of the companion star.

\subsubsection{The classical SFXT IGR~J17544-2619}

\begin{figure}
\includegraphics[width=7.cm]{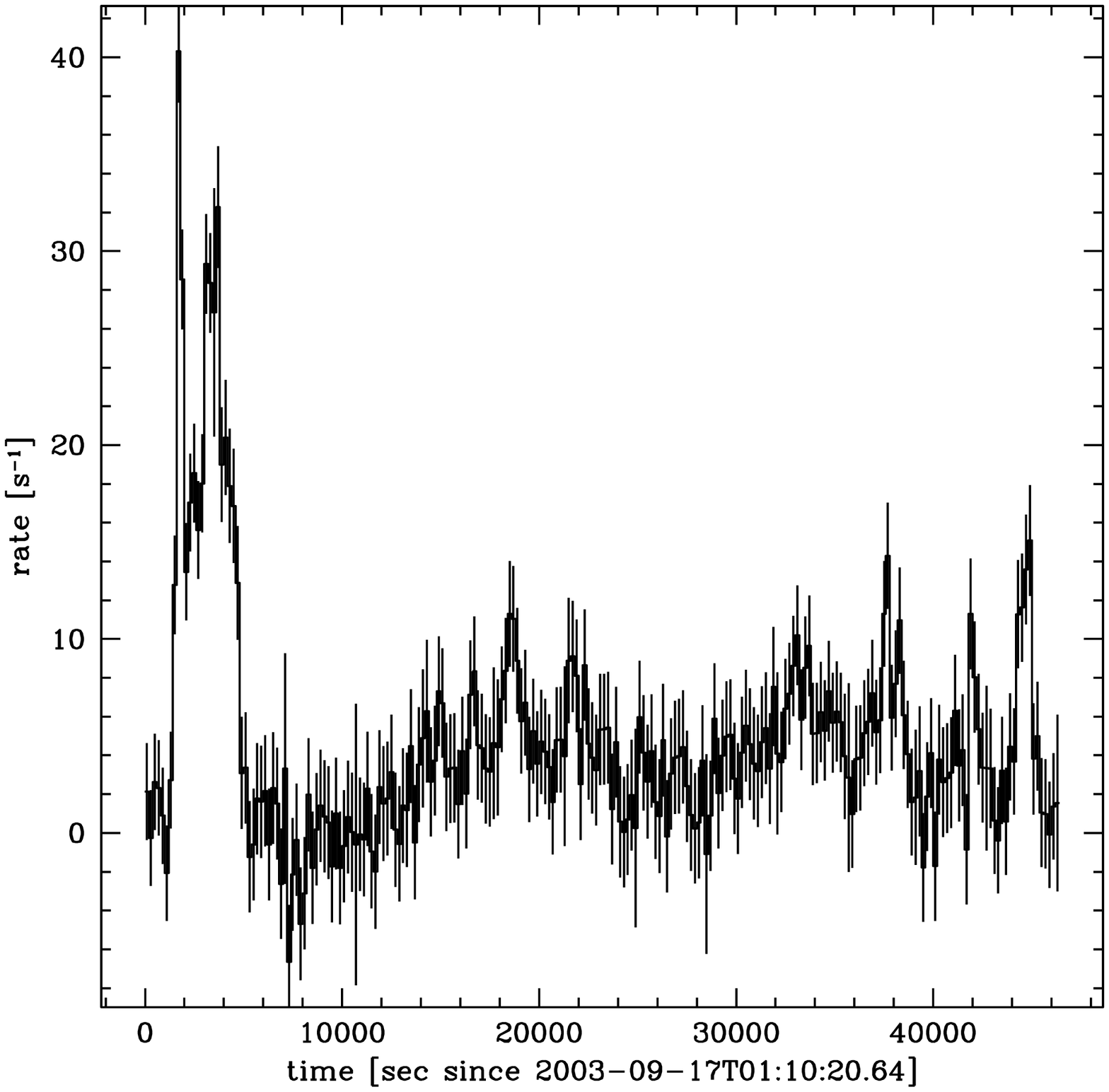}
\includegraphics[width=7.cm]{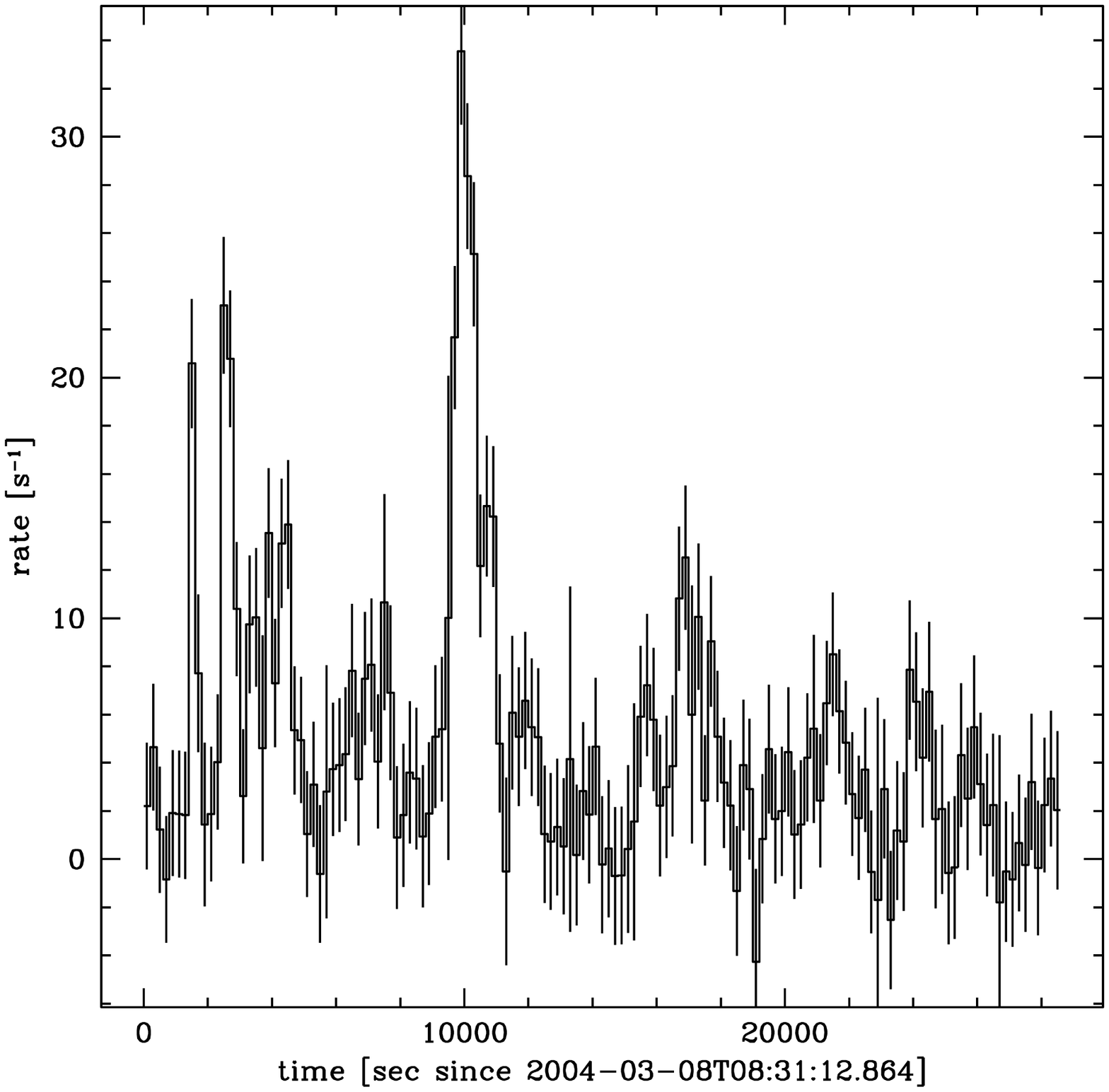}
\centerline{\includegraphics[width=11.cm]{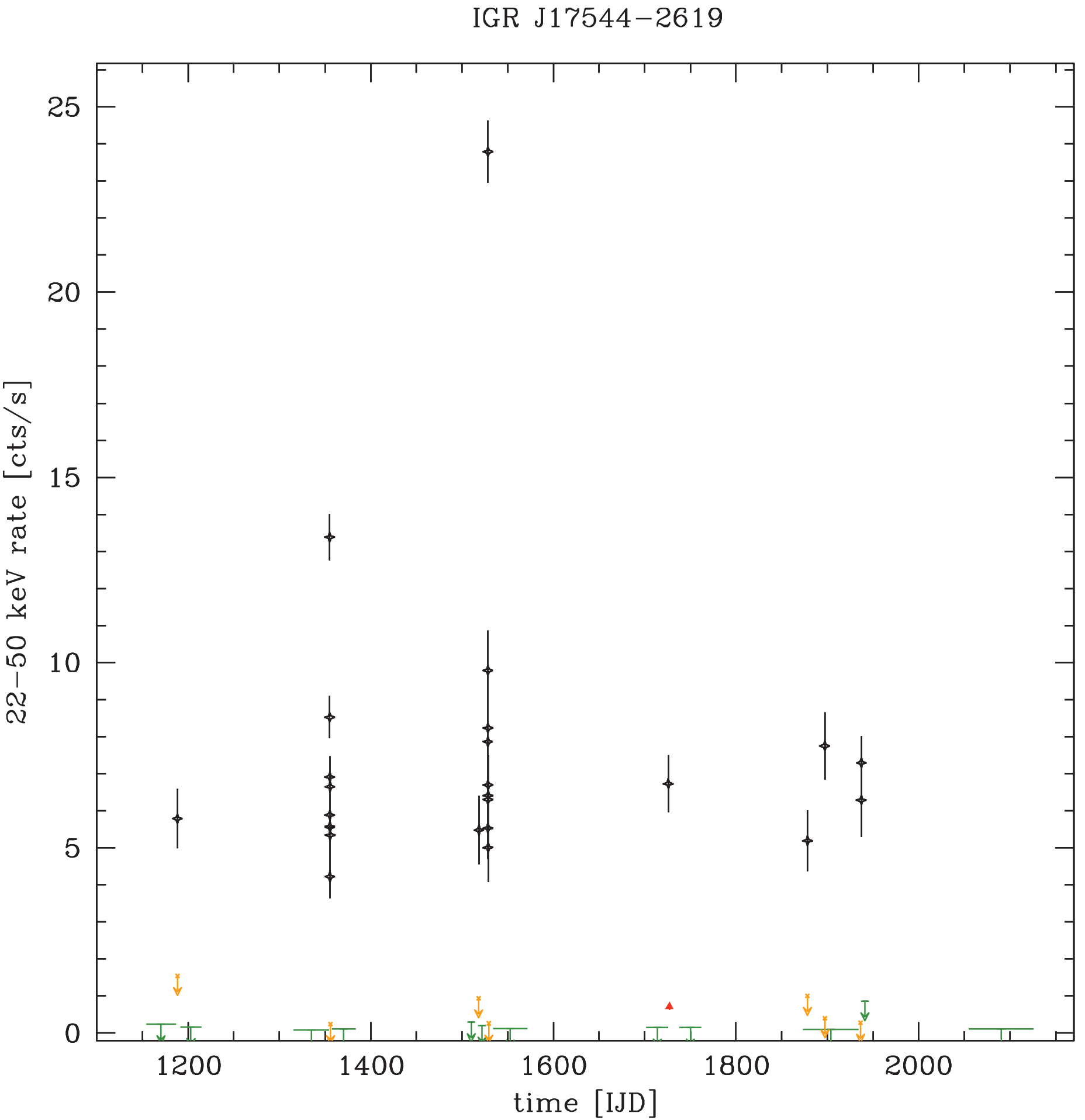}}
\caption{\label{figure:lcIGRJ17544}
{\it INTEGRAL} pointed (top panels, revolutions 113 and 171 respectively) and 200\,s binned (lower panel) light curves of IGR\,J17544-2619 (credit J.A. Zurita Heras).
}
\end{figure}

This bright recurrent transient X-ray source was discovered by {\it INTEGRAL} on 17 September 2003 \citep{sunyaev:2003b}. {\it XMM-Newton} observations showed that it exhibits a very hard X-ray spectrum, and a relatively low intrinsic absorption ($\nh \sim 2 \times 10^{22}\cmmoinsdeux$, \citeauthor{gonzalez-riestra:2004} \citeyear{gonzalez-riestra:2004}).  Its bursts last for hours, and inbetween bursts it exhibits long quiescent periods, which can reach more than 70\,days. The X-ray behaviour is complex on long, mean and short-term timescales: rapid flares are detected by {\it INTEGRAL} on all these timescales, on pointed and 200\,s binned light curve (see Figure \ref{figure:lcIGRJ17544}). The compact object is probably a NS \citep{intzand:2005}.  \citet{pellizza:2006} managed to get optical/NIR ToO observations only one day after the discovery of this source. They identified a likely counterpart inside the {\it XMM-Newton} error circle, confirmed by an accurate localization from {\it Chandra}.  Spectroscopy showed that the companion star was a blue supergiant of spectral type O9Ib, with a mass of $25-28 \Msol$, a temperature of $T\sim 31000$~K, and a stellar wind velocity of $265 \pm 20 \kms$ (which is faint for O stars): the system is therefore an HMXB \citep{pellizza:2006}. \citet{rahoui:2008a} combined optical, NIR and MIR observations and showed that they could accurately fit the observations with a stellar model of an O9Ib star, with a temperature $\Tstar \sim 31000$~K and a radius $\Rstar = 21.9 \Rsol$, deriving an absorption A$_v = 6.1 \mags$ and a distance D~$=3.6$~kpc. 


\subsubsection{The intermediate SFXT IGR\,J18483-0311} \label{section:IGRJ18483}

X-ray properties of this system were suggesting an SFXT nature
\citep{sguera:2007}, exhibiting however an unusual behaviour: its
outbursts last for a few days (to compare to hours for classical
SFXTs), and the ratio $L_{\mathrm max}/L_{\mathrm min}$ only reaches $\sim 10^3$ (compared to $\sim 10^4$ for classical SFXTs), therefore suggesting a high level quiescence. Moreover, $P_\mathrm{orb} = 18.5$\,days is low compared to classical SFXTs with large and eccentric orbits. Finally, $P_\mathrm{orb}$ and $P_\mathrm{spin} = 21.05$\,s values locate it inbetween Be and sgHMXBs in the Corbet diagram (see Figure \ref{corbet}).
\citet{rahoui:2008b} identified the companion star of this system
as a B0.5~Ia supergiant, unambiguously showing that this system is an
SFXT.  Furthermore, they suggested that this system could be the first
firmly identified intermediate SFXT, characterised by short, eccentric
orbit (with an eccentricity $e$ between 0.4 and 0.6),
and long outbursts: an intermediate SFXT nature would explain the unusual characteristics of this source among classical SFXTs.

\subsubsection{What is the origin of ``misplaced'' sgHMXBs?}

\citet{liu:2011} noted that there are two ``misplaced'' SFXTs in the Corbet diagram: IGR\,J11215-5952 \citep[a NS orbiting a B1 Ia star;][]{negueruela:2005b, sidoli:2007} and IGR\,J18483-0311 (described in the previous paragraph; see Figure \ref{corbet}). According to \citet{liu:2011}, these 2 SFXTs can not have evolved from normal main sequence OB-type stars, since {\it i.} their NS are not spinning at the equilibrium spin period $P_\mathrm{eq}$ of O\,V stars, {\it ii.} they have not been able to spin up after reaching $P_\mathrm{eq}$, and {\it iii.} their NS have not yet reached $P_\mathrm{eq}$ of sgHMXBs (see e.g. \citeauthor{waters:1989} \citeyear{waters:1989} and section \ref{section:corbet}). They must therefore be the descendants of BeHMXBs (i.e. hosting O-type emission line stars), after the NS has reached $P_\mathrm{eq}$, suggesting that some HMXBs exhibit two periods of accretion, the first one as BeHMXB and the second one as sgHMXB \citep{liu:2011}. This important result suggests {\it i.} an evolutionary link between BeHMXBs and sgHMXBs, and {\it ii.} that there must exist many more such intermediate SFXTs in the Be part of the Corbet diagram.


\subsection{A scenario towards the Grand Unification of sgHMXBs}

\begin{figure}
\includegraphics[height=.265\textheight,angle=0]{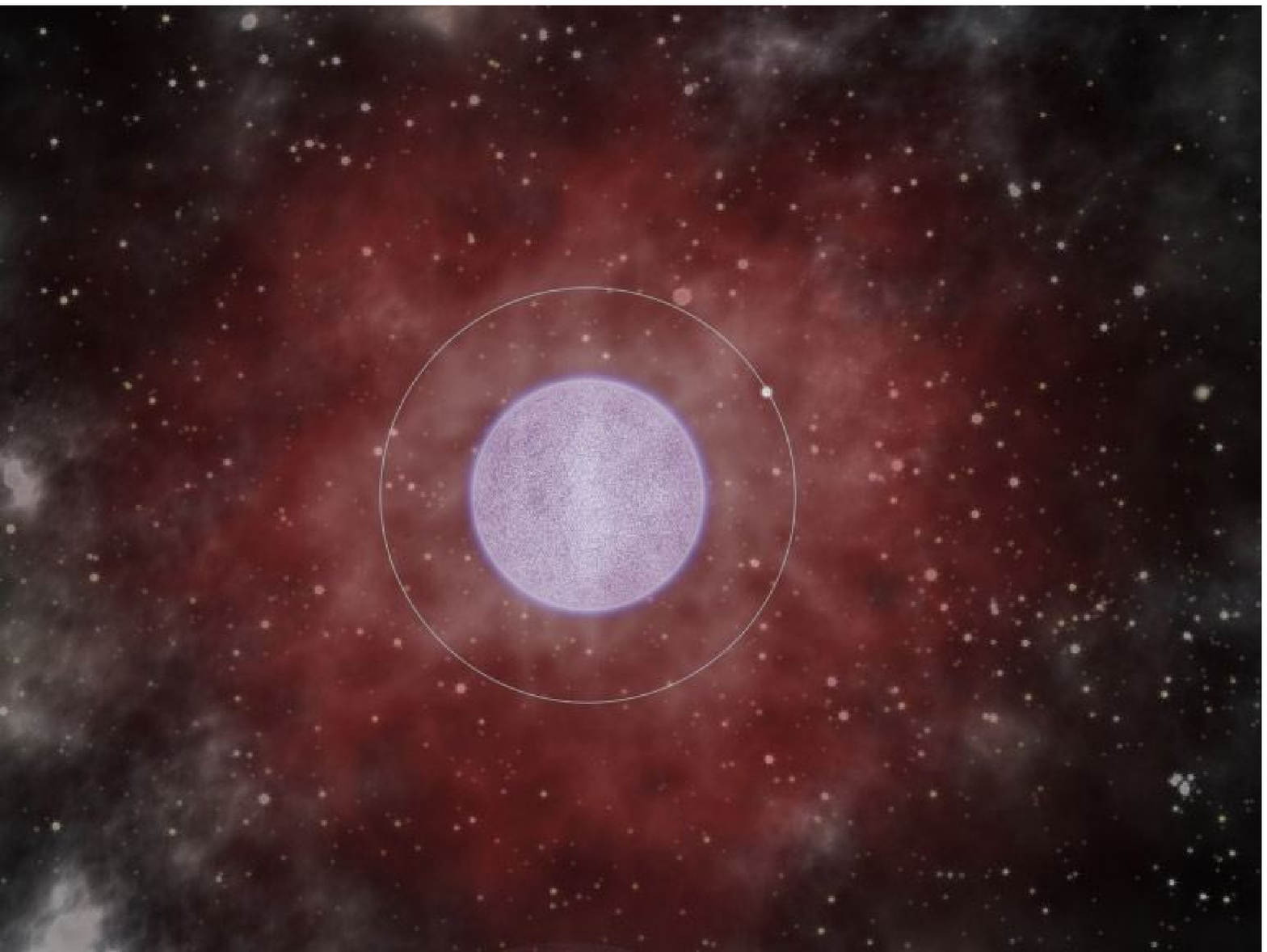}
\includegraphics[height=.265\textheight,angle=0]{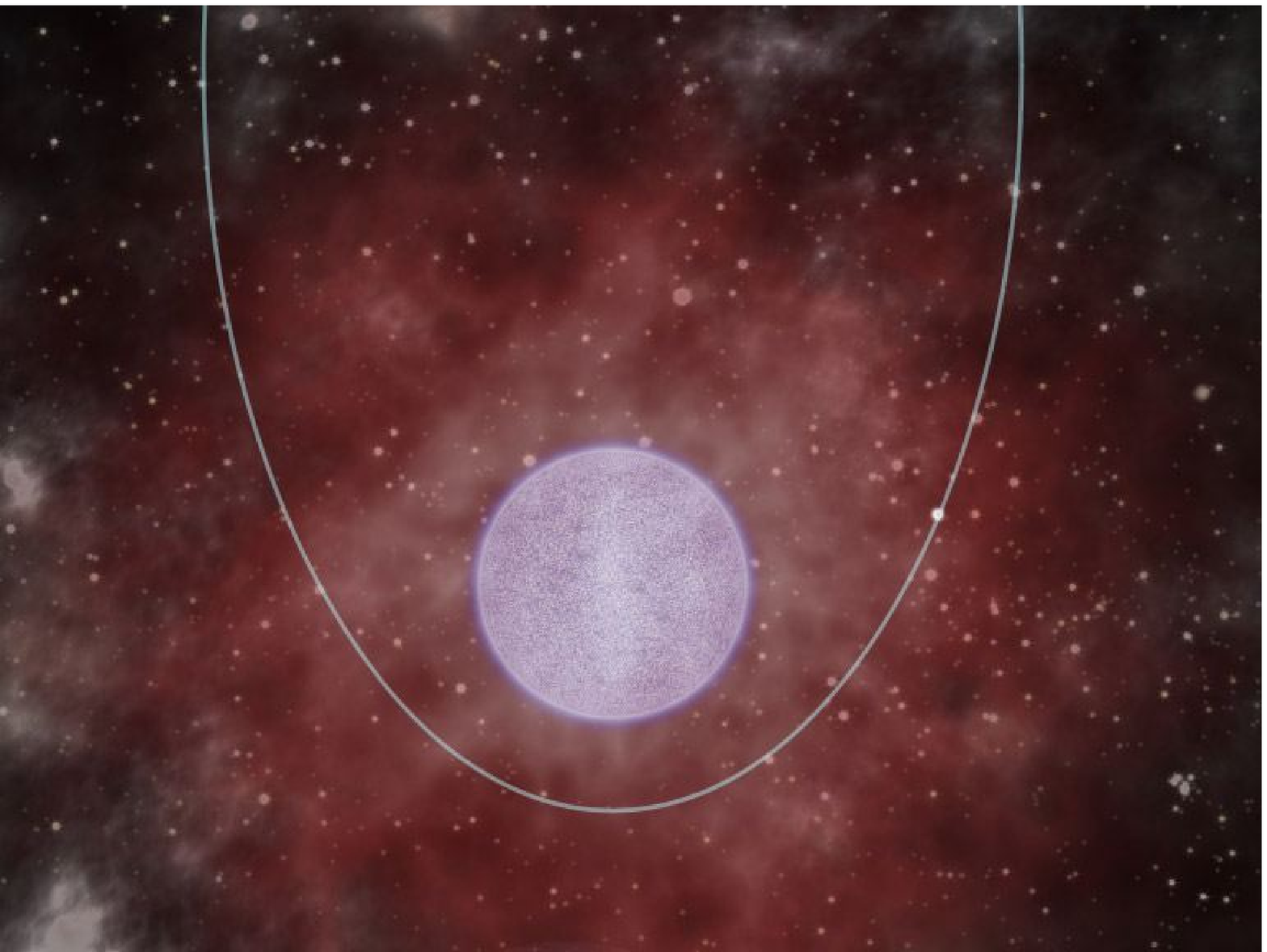}
\caption{\label{figure:obscured-sfxt}
  Scenario illustrating two possible configurations of {\it INTEGRAL} 
  sources: a NS orbiting a supergiant
  star on a circular orbit (left image); and on an eccentric orbit
  (right image), accreting from the clumpy stellar wind of the
  supergiant.  The accretion of matter is persistent in the case of
  the obscured sources, as in the left image, where the compact object
  orbits inside the cocoon of dust enshrouding the whole system. On
  the other hand, the accretion is intermittent in the case of SFXTs,
  which might correspond to a compact object on an eccentric orbit, as
  in the right image.  A 3D animation of these sources is available on the
  website: {\em http://www.aim.univ-paris7.fr/CHATY/Research/hidden.html}
}
\end{figure}

In view of the results described above, there seems to exist a continuous trend, with the observed division between classical sgHMBs and SFXTs which might be naturally explained by simple geometrical differences in the orbital configurations \citep[see e.g.][]{chaty:2008a}:

\begin{description}

\item [Classical (or obscured/absorbed) sgHMXBs:]
  These systems (like IGR J16318-4848) host a NS on  a short and circular orbit at a few stellar radii only from the star, inside the zone of stellar wind high clump density ($R_\mathrm{orb} \sim 2\Rstar$). It is constantly orbiting inside a cocoon of dust and/or cold gas, probably created by the companion star itself, therefore creating a persistent and luminous X-ray emission. The cocoon, with an extension of $\sim 10 \Rstar = 1$\,a.u., is enshrouding the whole binary system (see Figure \ref{figure:obscured-sfxt}, left panel).

\item [Intermediate SFXT systems:]
In these systems (such as IGR\,J18483-0311, $P_\mathrm{orb} = 18.5$\,days), the NS orbits on a short and circular/eccentric orbit outside the zone of high clump density, and it is only when the NS penetrates inside the narrow transition zone between high and low clump density, close to the supergiant star, that accretion takes place, and that X-ray emission arises, with possible periodic outbursts.

\item [Classical SFXTs:]
In these systems (such as XTE\,J1739-302, $P_\mathrm{orb} = 50$\,days), the NS orbits outside the high density zone, on a large and eccentric orbit around the supergiant star, and exhibits some recurrent and short transient X-ray flares, when it comes close to the star, and accretes from clumps of matter coming from the wind of the supergiant.  Because it is passing through more diluted medium, the $\frac{Lmax}{Lmin}$ ratio is higher and the quiescence lasts for longer time for classical SFXTs compared to intermediate SFXTs (see Figure \ref{figure:obscured-sfxt}, right panel).

\end{description}

Although this scenario seems to describe quite well the observational characteristics currently seen in most sgHMXBs, there are some for which it does not work: for instance the SFXTs IGR\,J16479-4514 and IGR\,J17544-2619 have orbital periods of $P_\mathrm{orb} = 3.32$\,days \citep{jain:2009} and $P_\mathrm{orb} = 4.926$\,days \citep{clark:2009} respectively, shorter than classical sgHMXBs. This shows that the macro-clumping scenario alone can not explain all properties of these sources, and that there must be additional mechanisms to take into account in order to explain these accretion processes, such as the formation of transient accretion discs \citep{ruffert:1996, ducci:2010} and the accretion with magneto-centrifugal barriers \citep{bozzo:2008}. Therefore, to better understand all mechanisms at work, we still need to identify the nature of more sgHMXBs to confirm it, and in particular $P_\mathrm{orb}$ and the dependance of the column density with the orbital phase of the binary system.

\section{Formation and evolution of High-Mass X-ray Binaries} \label{section:formation-evolution}

\subsection{Formation: Galactic distribution of X-ray binaries} \label{section:formation}

The evidence that the stellar birthplace is very important in the formation and evolution of X-ray binaries was initially noticed by examining the Galactic distribution of X-ray sources detected by {\it Ginga} \citep{koyama:1990} and {\it RXTE/ASM} \citep{grimm:2002}. Thereafter, a statistical analysis of a more complete sample of sources provided by {\it INTEGRAL} has been undertaken by \citet{lutovinov:2005a}, \citet{dean:2005}, \citet{bodaghee:2007} and \citet{bodaghee:2012b}. These works showed that LMXBs were concentrated towards the central regions of the Galaxy, more precisely in the Galactic bulge, and then gradually decreasing outside, while HMXBs, underabundant in the central kpc, were spread over the whole Galactic plane, confined in the disc, exhibiting an uneven distribution, preferentially towards the tangential directions of the spiral arms. 
This spatial distribution was expected, because LMXBs are constituted of companion stars belonging to an old stellar population, located in the Galactic bulge where they had time to migrate out of the Galactic plane ($|b| > 3-5 \adeg$). On the contrary, HMXBs -- containing young companion stars -- remain close to their stellar birthsite.

\begin{figure}
  \centerline{\includegraphics[width=15.cm]{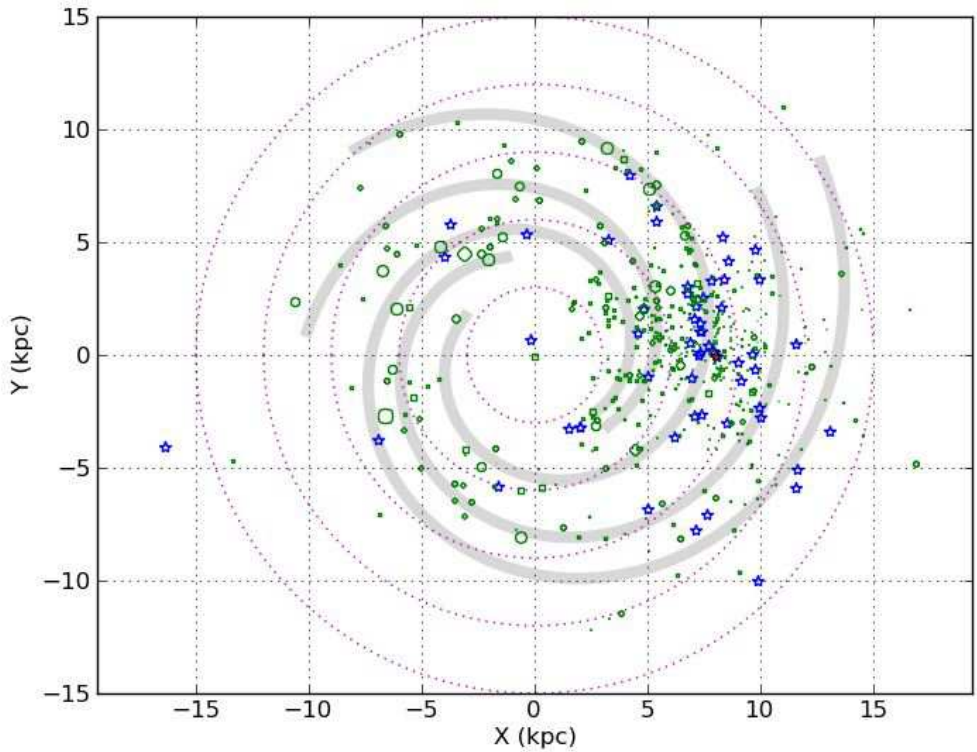}}
  \caption[Galactic distribution of HMXBs and SFCs]{Galactic distribution of HMXBs (blue stars) -- with distance derived by adjusting the spectral type on optical/infrared magnitudes of the companion star --, and of SFCs \citep[green circles, symbol size proportional to SFC activity;][]{russeil:2003}. The Sun is indicated by the red dot at (8.5, 0) (from \citeauthor{coleiro:2011} \citeyear{coleiro:2011}).
}
  \label{figure:distribution}
\end{figure}

To perform such a study, one has to derive the distribution of Galactic HMXBs from the catalogues of sources detected by current high energy satellites such as {\it XMM}, {\it Chandra}, {\it Swift} and {\it INTEGRAL}, taking into account the observational biases, some areas being more observed than others. However, the main difficulty is that the distance of most HMXBs is not accurately determined. To overcome this difficulty, a possibility is to determine the distance of all known HMXBs, by fitting their optical and infrared magnitudes with a black body corresponding to the companion star spectral type, and to compare their distribution with catalogues of massive stars, active OB stars and star formation complexes \citep[SFCs;][]{russeil:2003}. 
This method, undertaken and detailed in \cite{coleiro:2013a}, gives a typical clustering size of HMXBs with SFCs of $0.3 \pm 0.05$\,kpc, with an average distance between these clusters of $1.7 \pm 0.3$\,kpc, therefore clearly suggesting a correlation (see Figure \ref{figure:distribution}).

The appearance of HMXBs should theoretically follow the passage of the wave density associated with the rotation of the spiral arms \citep{lin:1969}, inducing a burst of stellar formation. In this context, the models predict a time lag between star formation and HMXB apparition, depending on the position in the Galaxy, and reflecting the range of initial masses of both stars composing each binary system \citep{dean:2005}. However, since the formation of an HMXB only takes tens of millions of years \citep{tauris:2006}, it is an accurate marker for the passing of density waves in a given region of our Galaxy. To produce a significant statistical study, we not only need many HMXBs for which we accurately know the distance, but also a precise and reliable kinematic model of the spiral arm structure of our Galaxy.
\cite{coleiro:2013a} investigate the expected offset between the position of spiral arms and HMXBs, and constrain the age and migration distance due to supernova kick for 13 sources, therefore assessing the influence of the environment on these high-energy objects with unprecedented reliability \cite[see][for more details]{coleiro:2013a}.

\subsection{Evolution: link with population synthesis models}

sgHMXBs revealed by {\it INTEGRAL} can allow us to better constrain the evolution of HMXBs, by comparing them to numerical study of population synthesis models. For instance, these new systems might represent a precursor stage of what is known as the "common envelope phase" in the evolution of HMXBs \citep{paczynski:1971}, when the orbit has shrunk so much that the NS begins to orbit inside the envelope of the supergiant star. However, the comparison between these new observations and population synthesis models is not straightforward, because many parameters influence the evolution of these massive compact binary systems in different ways, such as their mass, size, orbital period, age, rotation, magnetic field, accretion type, and stellar and circumstellar properties...

The formation and evolution of the {\it INTEGRAL} sgHMXBs is mentioned in the very nice review about relativistic binaries by \citet{vandenHeuvel:2009}. $P_\mathrm{orb}$ in later evolutionary phases being linearly dependent of initial $P_\mathrm{orb}$, we can derive that currently wide $P_\mathrm{orb} \sim 100$\,days OB-spectral type {\it INTEGRAL} sgHMXBs require initial systems with $P_\mathrm{orb} \sim 10$\,days \citep{vandenHeuvel:2009}.
Systems having long $P_\mathrm{orb}$ are expected to survive the common envelope phase, and may then either end as close eccentric binary radio pulsar systems (double NS), or in some cases as BH-NS binaries.

However, while some of these systems might potentially harbour a BH as the compact object, no purely wind-accretor system is formally known yet. Of course NSs are easier to detect through X-ray pulsations, but, as Carl Sagan pointed it out, {\it ``absence of evidence is not evidence of absence''}. We should look for BHs orbiting supergiant companion stars in wind-accreting HMXBs, however this is only feasible through observational methods involving detection of extremely faint radial velocity displacement due to the high-mass of the companion star, or through extremely accurate radio measurements that will be available in the future. On the other hand, massive stars lose so much matter during their evolution that they might always end up as NSs (see e.g. \citeauthor{maeder:2008} \citeyear{maeder:2008}). If this is the case, then such systems hosting BHs might not form at all. 

Finally, these sources are also useful to look for massive stellar progenitors, eventually evolving in Luminous Blue Variables (LBVs) or Wolf-Rayet (WR) stars, and finally giving birth to coalescence of compact objects, through NS/NS or NS/BH collisions. They would then become prime candidates for gravitational wave emitters, or even short-hard $\gamma$-ray bursts.

\section{Conclusions and perspectives} \label{section:conclusion}

While the {\it INTEGRAL} satellite was not primarily designed for this, it allowed a great progress in the understanding of HMXBs in general, and of sgHMXBs in particular. Let us first recall the {\it INTEGRAL} legacy concerning sgHMXBs:

\begin{itemize}

\item The {\it INTEGRAL} satellite has quintupled the total number of known Galactic sgHMXBs, constituted of a NS orbiting a supergiant star. Most of the new sources are slow and absorbed X-ray pulsars, exhibiting a large $\nh$ and long $P_\mathrm{spin}$ ($\sim1$\,ks), typical of wind-fed accreting pulsars (according for instance to the Corbet diagram).

\item The {\it INTEGRAL} satellite has revealed the existence in our Galaxy of two previously hidden populations of high energy binary systems.
First, a population of obscured and persistent sgHMXBs, composed of supergiant companion stars exhibiting a strong intrinsic absorption and long $P_\mathrm{spin}$, with the NS deeply embedded in the dense stellar wind, forming a dust cocoon enshrouding the whole binary system.
Second, the SFXTs, exhibiting brief and intense X-ray flares -- with a {\bf luminosity $L_X \sim 10^{36} \ergs$ at the peak, during a few ks} every $\sim 7$\,days --, which can be explained by accretion through clumpy winds.

\end{itemize}

However, one may ask the questions: Apart from these observational facts, has the {\it INTEGRAL} satellite allowed us to better understand all populations of HMXBs, by increasing their number, compared to the ones already known before its launch? 
Do we better apprehend the accretion processes in HMXBs in general, and in sgHMXBs in particular, and what makes the fast transient flares so special, in the context of the clumpy wind model, of the formation of transient accretion discs, and/or the centrifugal/magnetic barrier? 

The answers to these questions is probably {\it ``not yet''}, however we now have in hand more sources, and therefore more constraints to play with.
Studying these populations will provide a better understanding of the formation and evolution of short-living HMXBs, and study accretion processes. Finally, it is clear that stellar population synthesis models now have to take these objects into account, to assess a realistic number of all populations of high energy binary systems in our Galaxy. 






\end{document}